\documentclass[12pt]{iopart}

\usepackage{color}

\usepackage{graphicx}

\begin{document}

\title[Controlling protein crystal growth rate by means of temperature]{Controlling protein crystal growth rate by means of temperature}

\author{I. Santamar\'ia-Holek$^1$\footnote{Permanent address: UMJ-Facultad de Ciencias, Universidad Nacional Aut\'onoma de M\'exico,
Boulevard Juriquilla 3001, Juriquilla 76230, Quer\'etaro, M\'exico}, Adam Gadomski$^1$ \& J. M. Rub{\'i}$^2$}

\address{$^1$Institute of Mathematics \& Physics, University of Technology \& Life Sciences, 85796 Bydgoszcz, Poland
\newline $^2$Departament de F{\'i}sica Fonamental,
University of Barcelona, 08028 Barcelona, Spain}
\ead{$^{1a}$agad@utp.edu.pl; $^{1b}$isholek.fc@gmail.com $^2$mrubi@ub.edu}
\begin{abstract}
We have proposed a model to analyze the growth kinetics of lysozyme crystals/aggregates under non-isothermal conditions. The model was formulated through an analysis of the entropy production of the growth process which was obtained by taking into account the explicit dependence of the free energy on the temperature. We found that the growth process is coupled with temperature variations resulting in a novel Soret-type effect. 
We identified the surface entropy of the crystal/aggregate as a decisive ingredient controlling the behavior of the average growth rate as a function of temperature. The behavior of the Gibbs free energy as a function of temperature is also analyzed. The agreement between theory and experiments is very good in the range of temperatures considered.
\end{abstract}

\maketitle

\section{Introduction}
Experimental and theoretical studies on the active role of temperature in protein crystal 
formation and other related problems of soft-matter aggregation could open a relevant 
possibility toward controlling the morphological and structural properties of protein crystals 
through an adequate tunning of the thermal conditions in which they grow \cite{pusey1,pusey2,ducrg,Bartling,glatt,lizoz}. 
The importance of this research becomes manifest in biomedical \cite{Bartling} and clinical studies 
where the effect of temperature on forming certain (disease-assisted) 
plagues or fibrils was observed to help in partitioning aggregates into several groups of 
pathomorphological interests \cite{clinic}. 

The research on the active role of temperature in exploring optimal nucleation and growth 
conditions of protein crystals or aggregates is mostly empirical. A large 
number of experiments conducted in thermostatted conditions had
contributed to  the comprehension of the sen\-si\-ti\-ve-to-tem\-pe\-ra\-tu\-re kinetics 
by monitoring the temperature dependence of related thermodynamic quantities such as the growth rate of the crystals/aggregates \cite{pusey1,pusey2,ducrg,Bartling,glatt,lizoz}.
For example, in recent years the effect of temperature on average $(110)$ face growth rates in the range of $277-293 K$ at different aqueous solution and salt concentrations has been measured carefully \cite{pusey}.  This experimental study has shown that the crystal formation appears ultimately to be a common product of interfacial and solution involving mechanisms. The formation, seen in terms of 
interface dynamics, involves the existence of the so-called depletion zone \cite{pusey1}, 
that is, a screened interfacial force field of electrostatic and hydrophobic nature whose characteristic length is temperature dependent \cite{ducrg,lizoz,pusey,navarro,pvekil}. 

Here, we present a theoretical model accounting for the average growth rate 
of crystals/aggregates as a function of temperature. We start by calculating the 
mesoscopic entropy production \cite{rubi} of the growth process that contains 
contributions associated to two variables, the radius and 
the temperature of the crystal/aggregate. These contributions lead to a 
\textit{novel} Soret-type effect coupling the dynamics of the radius 
with the variations of the temperature, in a way similar to that of the conventional
thermal diffusion
\cite{old,dean,brown2,wieg2}. After explicitly considering the dependence of the 
Gibbs free energy on temperature through the surface energy term, a formula is
obtained and used to make quantitative comparisons with experiments. 
By discriminating clearly the effects of the surface energy and temperature, our model 
may also help in discerning the degree of importance of the electrostatic 
screening effect and of temperature variations on the appearance of different surface 
morphologies and therefore of the output forms \cite{Bartling,putn,rolando,agg3,agg4,andrej,kula}.  

The proposed model may constitute a comprehensive and useful theoretical
framework for describing the formation of lysozyme crystals/aggregates 
and its growth in temperature-controlled conditions by addressing:  
\textit{a)} The free-energy behavior as a function of temperature in lysozyme 
crystallization, and \textit{b)} how this behavior is influenced by the temperature
through the contribution of the surface entropy of the crystals/aggregates. 
Finally, \textit{c)} we formulate a quantitative 
model accounting for the behavior of the crystal growth rate as a function
of temperature that takes into account a coupling term associated to the
thermal expansion of the system and mediated by the Soret-type effect. 
The results of the model show good agreement with experiments.

The paper is structured as follows. In Sec. 2 we calculate the mesoscopic entropy production
leading to the Soret-type diffusion in cluster size $R$-space.
Then, in Sec. 3 we discuss the dependence of the Gibbs free energy
on temperature and its influence in crystal's surface energy and entropy, 
and show how the Soret-type coefficient and the average growth rate of
crystals/aggregates emerge from our analysis.
Sec. 4 is devoted to formulate the model for the average growth rate and to compare
with experiments. In the final section, we shall conclude on the main findings and their inherent effect on the model soft-matter aggregating system.

\section{Entropy production in protein crystal formations: Size and temperature}
The characterization of the nonequilibrium 
state of the protein 
aggregate is performed through the knowledge of the probability 
density $P \equiv P(R,t),$  of finding the system with an effective radius 
$R\in[R,R+dR]$ at time $t$. This quantity obeys the continuity equation
\begin{equation}
\frac{\partial }{\partial t}P(R,t)+\frac{\partial}{\partial R}
J(R,t) = 0,
\label{continuity1}
\end{equation}
where $J(R,t)$ is an unknown probability current in $R$-space. Protein aggregation
is an irreversible process and as such is governed by the thermodynamics laws. To
determine the value of $J(R,t)$, we will make use of the Gibbs' entropy
postulate in the form \cite{rubi}
\begin{equation}
S(t) = S_{leq}- k_{B}\int P\ln\frac{P}{P_{leq}}dR ,
\label{entropy postulate}
\end{equation}
where $S(t)$ and $S_{leq}$ are the nonequilibrium and local equilibrium entropies
of the system. The equilibrium value of the probability $P_{leq}({R})$ 
is given by the canonical distribution function \cite{rubi,agg1,landau}  
\begin{equation}
P_{leq}(R)=\frac{1}{Z}\e^{-W_{min}/k_BT}. \label{peq general}
\end{equation}
Here, $Z\equiv Z(T)$ is the canonical partition function that depends on 
the temperature of the aggregate $T$ which is imposed by a heat bath.
$W_{min}$ is the minimum reversible work required to create the actual state 
of the aggregate and $k_B$ is the Boltzmann constant. It is important 
to notice here that the radius $R$ of the aggregate is in general a function of
the temperature, $R=R(T)$. This fact will be taken into account in what follows.

At constant pressure and temperature, the minimum-work associated to the formation 
of the protein aggregate can be identified with a free energy difference \cite{rubi}
\begin{equation}
W_{min}(R,T)=\Delta G(R,T). \label{minimum work}
\end{equation}

Protein aggregation entails rearrangements of its intrinsic structure, a
process which is only possible at a certain energetic cost. The rate at
which energy is dissipated is $T \sigma$, where $\sigma$ is the entropy
production rate \cite{rubi}. This quantity follows from the time
derivative of Eq. (\ref{entropy postulate}) after taking into account that in the
local equilibrium state
\begin{equation}
\frac{\partial S_{leq} (t) }{\partial t} = T^{-1}\frac {\partial E(t)}{\partial t}
= - T^{-1} \int \frac{\partial J_q}{\partial R} \, dR
\end{equation}
where $J_q(R,t)$ is the amount of heat exchanged per unit time 
during the growth of the crystal/aggregate \cite{deGroot}. An
integration by parts after employing (\ref{continuity1}) yields an evolution equation for the
local entropy $s(R,t)$ of the form:
${d s}/{d t} + {\partial J_s}/{\partial R}=\sigma$.
Here, $J_s$ is the entropy current and the
entropy production rate $\sigma(R,t)$ is given by \cite{mazur1,mazur2}
\begin{equation}
\sigma(R,t) = -\frac{\tilde{J_q}}{T^2} \frac{\partial T }{\partial R}
-\frac{J}{T} \left[ \left(\frac{\partial \Delta G}{\partial R}\right)_{T} + \frac{k_BT}{ P}  \frac{\partial P}{\partial R} \right].
\label{entrop-prod}
\end{equation}
This equation establishes that energy dissipation during the evolution in time of the system is 
due to two contributions. The first one comes from the dependence of the radius of 
the aggregate on the temperature. The second contribution comes from the driving 
force leading the aggregation in $R$-space, and coming from the explicit dependence of the 
nonequilibrium chemical potential $\mu(R,T,t)$ on the radius of the crystal/aggregate.  
Here, $\mu(R,T,t)$ is defined by 
\begin{equation}
\mu(R,T,t)\equiv\mu_{leq}(T)+ k_B T \ln\left[P/P_{leq}\right] ,
\label{chem pot}
\end{equation}
where $\mu_{leq}(T)=-k_B T \ln Z(T)$. In deriving Eq. (\ref{entrop-prod}) we have 
introduced the modified heat current $\tilde{J_q}=J_q-h_f J$ that incorporates the 
enthalpy of formation of the aggregate $h_f$ and we have separated the variation of 
the chemical potential in the form:
$d\mu= \left( {\partial \mu}/{\partial R}\right)_{T,P}  dR+\left( {\partial \mu}/{\partial P}\right)_{T,R} dP+ \left( {\partial \mu}/{\partial T}\right)_{R,P} dT$. In this relation the subindexes  indicate the variables that should be considered constant when performing the derivative \cite{mazur1}.

From the entropy production (\ref{entrop-prod}) it is possible to formulate the following linear law for $J(R,t)$:
\begin{eqnarray}
J(R,t) =  -\gamma \left[ P \left(\frac{\partial \Delta G}{\partial R}\right)_{T}
+ k_BT \frac{\partial P}{\partial R} \right] -D_T P \frac{\partial T }{\partial R} ,
\label{linear-laws1}
\end{eqnarray}
where $\gamma$ and $D_T$ are, respectively, the mobility and
thermal diffusion-type coefficients
in the space of crystal/aggregate radius $R$.
After substituting Eq. (\ref{linear-laws1}) into (\ref{continuity1}) we
obtain  a Fokker-Planck type equation
\begin{equation}
\frac{\partial }{\partial t}P(R,t)=
\frac{\partial}{\partial R}\left[\gamma  \left(\frac{\partial \Delta G}{\partial R} \right)_{T}P
+ {k_BT} \gamma  \frac{\partial P}{\partial R} \right] +\frac{\partial}{\partial R}\left( D_T P \frac{\partial T }{\partial R}\right) .
\label{FP}
\end{equation}
This equation constitutes the starting point in the formulation of our model accounting for the dependence on temperature of the growth rate of protein-type aggregations as, for instance, lysozyme aggregations in aqueous solutions \cite{pozn,dren,poon}. This dependence has been reported in
experiments that confirm the sensitivity of the growth of the aggregates to temperature conditions \cite{pusey1,pusey2,ducrg,Bartling,pusey,navarro}, and can be analyzed through an adequate choice of the Gibbs energy and the specific physical properties
of each protein crystal/aggregate such as thermal expansion coefficient and surface energy.

\section{Temperature dependence of the Gibbs energy and the Soret-type effect}
In order to quantify the effect of temperature in protein aggregations 
it is convenient
to first derive the evolution equation for the average radius of the aggregate 
$R(t)$, defined by the relation: $R(t)=\int R\,P\,dR$.

Thus, according to its definition, the expression for the average growth rate of the (spheroidal) aggregate $V_a=d R(t) /dt$
is obtained by multiplying Eq. (\ref{FP}) by $R$ and then integrating
by parts over $R$-space.
Assuming that fluctuations are small, then the probability can be approximated by a
Dirac delta function of the form: $P \sim \delta (R(t)- R)$. One obtains
\begin{equation}
\frac{d R(t)}{dt}
= - D \left\{ \frac{1}{k_BT} \left[\frac{\partial }{\partial R}\Delta G(R(t),T)\right]_T + S_T \frac{\partial T }{\partial R} \right\} ,
\label{R-dot}
\end{equation}
where we have defined $D\equiv k_BT \gamma$ and the Soret-type coefficient
$S_T \equiv D_T/D$, in a way fairly reminiscent of that presented in \cite{wieg2,putn} 
for conventional thermodiffusion. Eq. (\ref{R-dot}) shows that under non-isothermal conditions, the growth rate of the aggregate depends on the diffusion constant and free energy barrier, as occurs in the isothermal case,  and on local variations of the temperature along the radius of the crystal/aggregate. This new contribution results from the coupling between the growth kinetics with the temperature inhomogeneities that define a Soret-type effect in R-space.
It is opportune to note that $V_a$, as an averaged measure 
proposed to quantify the growth rate, may differ from the instantaneous growth rate 
proposed earlier \cite{agg1,agg3} for the aggregations exploited as isothermal formations 
\cite{agg2,jcp10}.

At equilibrium ($V_a=0$) the Soret-type coefficient is determined from
Eq. (\ref{R-dot}) by the dependence of the free energy on the temperature in the form
\begin{equation}
S_T = - \frac{1}{k_BT}\left[ \frac{\partial }{\partial T}\Delta G(T)\right]_{R} .
\label{ST-Req}
\end{equation}
This relation is similar to that reported in the literature for the Soret effect
examined in particles' position space \cite{dean,brown2}.

In this way, the explicit dependence of $S_T$ on the temperature 
can be obtained by assuming that the free energy change is given by the expression of the classical nucleation theory
\begin{equation} \label{GCNT}
\Delta G=-({4\pi}/{3 v_p}) R^3\Delta \tilde{\mu} + 4 \pi \Gamma R^2,
\end{equation}
where $\Delta \tilde{\mu}(T)=\tilde{\mu}_c-\tilde{\mu}_i$ is the electrochemical supersaturation given by the electrochemical potential difference between the crystal $\tilde{\mu}_c$ and the protein solution $\tilde{\mu}_i$. Here, $\Gamma$ is the surface 
energy. 
This expression assumes the presence of a collective interfacial tension mechanism for the aggregation, originally proposed in Ref. \cite{iaco} for some inhomogeneous protein (lysozyme) solutions. Moreover,  we have tacitly supposed that in first approximation the 
form of the protein can be assimilated to an effective sphere with a gyration radius.  Recent results on computer simulations of the structure of certain proteins under non-isothermal conditions \cite{bresmepccp} have shown that the fluctuations of their radius of gyration are small thus indicating the validity of such an effective model. The symbol $v_p$ thus stands for the protein effective volume and the number of proteins in the aggregate can be obtained through the relationship $n_a=4\pi R^3/3 v_p$.

The temperature dependence of the physical parameters entering Eq. (\ref{GCNT}) can
be inferred on the following basis. According to the classical nucleation theory, the chemical potential difference in Eq. (\ref{GCNT}) is customarily assumed to be of the form: $\Delta\tilde{\mu}(T)=kT \ln|{\cal C }/c_i|$, with $c_i$ the density of proteins in solution and ${\cal C }$ the density of the crystal/aggregate. In this approximation,
$\ln|{\cal C }/c_i|$ can be considered as independent of $R$ and $T$ within the range of temperatures of interest for the (dis)orderly aggregation such as the one of lysozyme type in aqueous solution, taking mostly place at $pH \approx 4.5$ and up to about $T \simeq 310 K$ (high-temperature aggregation) \cite{agg4,agg1}. In addition, one has to consider the dependence of the protein volume $v_p$ on the temperature. According to Ref. \cite{truskett1}, this dependence is negligible for proteins in their native conformations and relatively weak for denatured proteins over the range of temperatures of interest. For convenience, in what follows we will assume that the protein volume is constant.

As a consequence of the previous considerations, we may assume that the 
only parameter depending explicitly on the temperature in Eq. (\ref{GCNT}) is the 
surface energy $\Gamma (T)$. It may be expressed as a power expansion of the 
temperature with respect to a characteristic temperature $T_0$. 
Up to second order one obtains
\begin{equation}\label{GammadeT-expansion}
\Gamma (T) \simeq \Gamma_0 + \left(\frac{\partial \Gamma}{\partial T} \right)_p (T-T_0) 
+ \frac{1}{2}\left(\frac{\partial^2 \Gamma }{\partial T^2} \right)_p (T-T_0)^2,
\end{equation}
where $\Gamma_0$ is the value of the surface energy at $T=T_0$. 
This relation can be written in terms of the surface entropy \cite{ragone}:  
$S_s \equiv - (\partial \Gamma /\partial T )_p$. Substituting 
the previous relation into (\ref{GammadeT-expansion}) we finally obtain
\begin{equation}\label{GammadeT}
\Gamma (T) \simeq \Gamma_0 -S^{(s)}_{0}(T-T_0) + b (T-T_0)^2,
\end{equation}
where  
$b=(1/2)(\partial^2 \Gamma /\partial T^2 )_p $.
Here, the characteristic value of the surface entropy at $T=T_0$ is assumed to 
be positive: $S^{(s)}_{0}>0$. Concerning the value of the parameter $b$ 
two possibilities can be analyzed \cite{dren}: 
\textit{i}) For $b < 0$, the free energy is a monotonic decreasing function of $T$.
\textit{ii}) For $b > 0$, after a first linear decrease which is dominated by the
volumetric and surface entropy terms, the free energy presents a minimum due to the quadratic contribution.
This second possibility reproduces the expected behavior of the free energy as 
a function of the temperature \cite{truskett1}, 
as shown in Fig. 1 obtained with the use of Eqs. (\ref{GammadeT}) and (\ref{GCNT}). 
The appearance of this minimum makes it possible the control of the growth dynamics, as will be shown in the analysis of the behavior of the growth rate.
Figure 2 shows in turn the Soret-type coefficient (\ref{ST-Req}) as
a function of the temperature. It is interesting to mention that this
coefficient exhibits a change of sign similar to that observed in the conventional Soret effect \cite{old,dean,brown2,wieg2}.
The four cases shown in Figs. 1 and 2 correspond to the 
values of the parameters obtained from the fit shown in Fig. \ref{V} and given in Table 
\ref{table1} and in the text.  The value of $T_0$ in each case corresponds to the initial point of the
growth rate in Fig. \ref{V}.

\begin{figure}[tbp]
\mbox{\resizebox*{15.0cm}{7.0cm}{\includegraphics{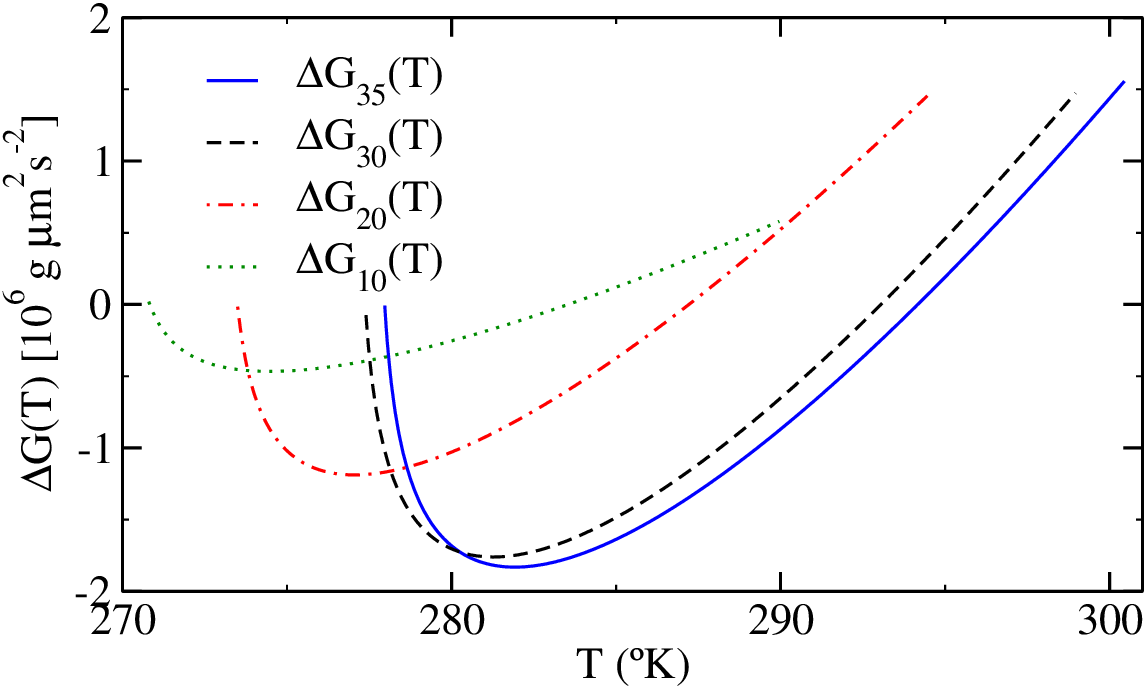}}} {}
\caption{The free energy given by Eq. (\ref{GCNT}) as a function of the temperature
in the case when the surface energy $\Gamma (T)$ is given by Eq. (\ref{GammadeT}). The free energy has the expected behavior, cf. Ref. \cite{truskett1}. The values of the parameters used are those obtained
in the fit shown in Fig. \ref{V}.}
\label{f}
\end{figure}

Here, it is worth mentioning that for the long stage isothermal growth of the crystal/aggregate, it was proposed in Refs. \cite{pvekil,agg1,agg2,jcp10} that the free energy $\Delta G(R,T)$ can be modeled through the near-interface electrochemical supersaturation of the system by $\Delta G(R,T) =  {k_BT} \ln |\delta_{i}|$,
where $\delta_{i}\equiv \delta_{i} (R,T)= {c_{i}(R,T)/[{\cal C}-c_{i}(R,T)]}$ is the inverse near-interface solubility $\Sigma_{i} (R,T) = {1/\delta_{i}} $. This quantity is expected to be measurable in terms of the inverse of the corresponding depletion-zone supersaturation. 
Here ${\cal C}$ is again the density of the aggregate and now $c_i(R,T)$ is the near interface
concentration of proteins, whose expression is given in turn by
$c_{i} (R,T) = c_0[{1+(2\Gamma/R)(1/k_BT)} ]$.
Here, $c_0$ is the concentration of proteins in the solution upon flatness of the
aggregate surface and $c_{i} (R,T)$ is viewed as the near interface density that incorporates a
dependence on the curvature and temperature. Notice that now $\Gamma/k_BT $ becomes
the Gibbs-Thomson capillary length \cite{agg1}. The dependence of $c_{i} (R,T)$ on $T$
introduces a correction of the supersaturation making it of non-ideal nature. 

This protein aggregation model makes it possible to incorporate second
and higher order contributions of a virial expansion
of the local chemical potential. The dependence of the second virial coefficient on temperature
has been analyzed in Ref. \cite{pvekil}. Finally, it is worth to mention that the generalized
form of the near interface concentration $c_{i} (R,T)$ is consistent with the Gibbs-Thomson boundary condition coming from Eq. (\ref{GCNT}) when $\Delta G$ is minimized.

\section{Comparison with experiments}
From the previous analysis, one may conclude that in non-isothermal experiments
performed at constant supersaturations \cite{pusey2,pusey}, the Soret-type effect 
characterized by the coefficient $S_T$ can be relevant in determining
the dependence of the growth rate $V_a$ on the temperature.
\begin{figure}[tbp]
\mbox{\resizebox*{15.0cm}{7.0cm}{\includegraphics{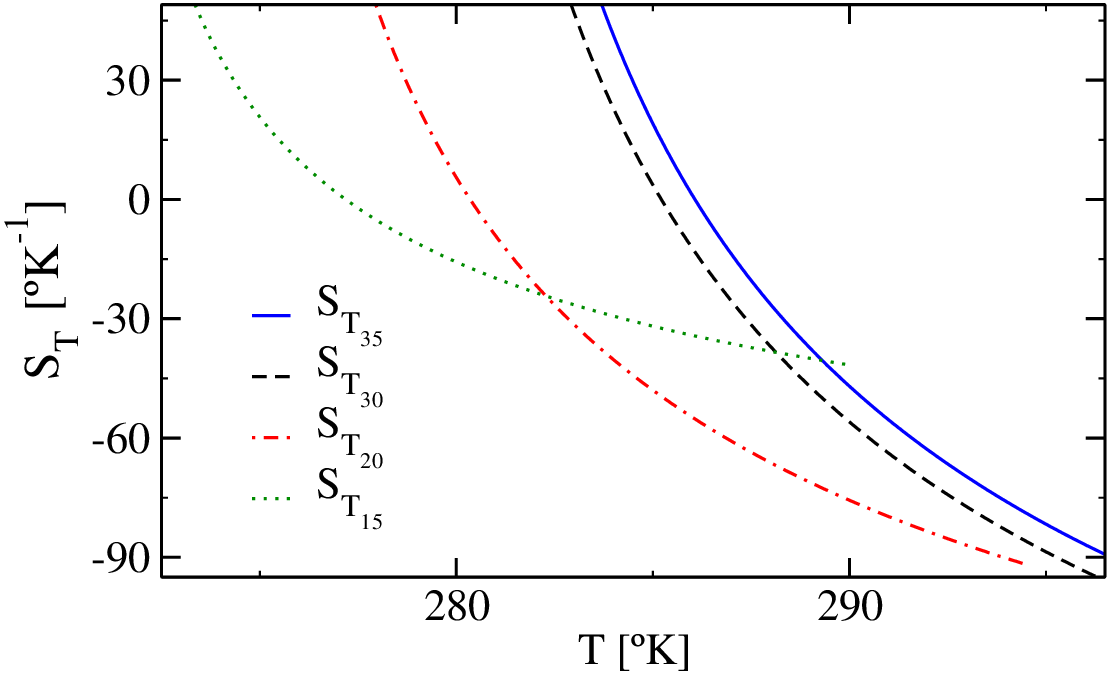}}} {}
\caption{ The Soret-type coefficient $S_T$ given by
Eq. (\ref{ST-Req}) combined with (\ref{GCNT}) and (\ref{GammadeT}) as a function of
the temperature and corresponding to the free energy of Figure 1. The values of
the parameters were those obtained from the fit in Fig. \ref{V}.}
\label{St}
\end{figure}

\begin{figure}[tbp]
\mbox{\resizebox*{15.0cm}{7.0cm}{\includegraphics{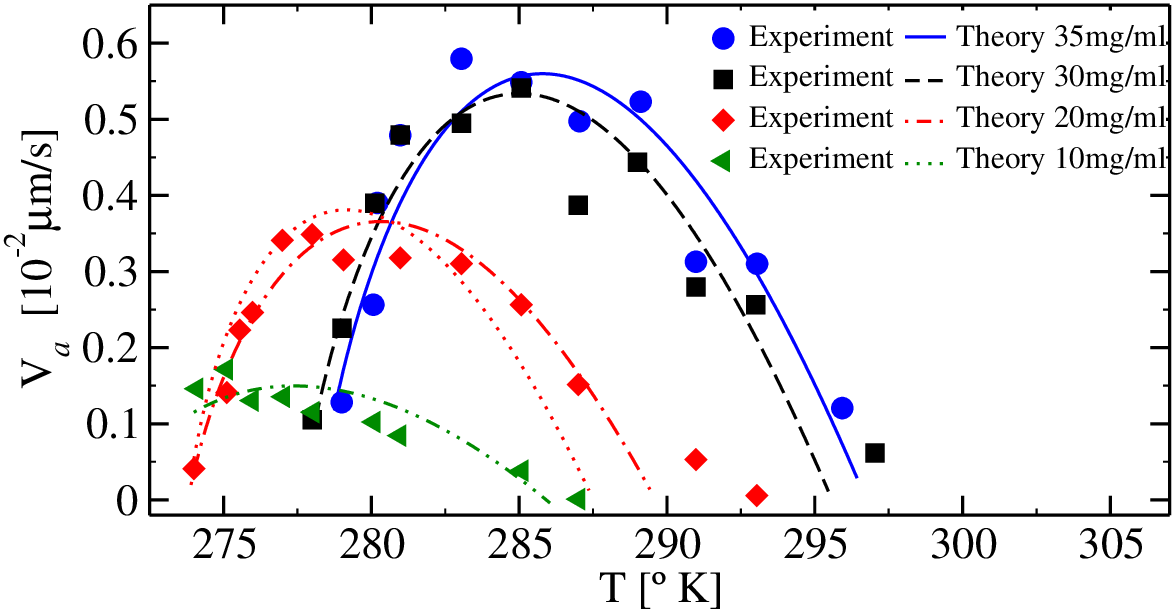}}} {}
\caption{Average growth rate in terms of temperature
($^oK$) for four different values of the supersaturation corresponding to protein
concentrations of $35mg/ml$ (solid line), $30mg/ml$ (dashed line), $20mg/ml$ 
(dash-dotted line) and $10mg/ml$ (dash double-dotted line). The 
symbols are experimental results taken from \cite{pusey2} whereas the lines 
correspond to our proposed fitting model, Eq. (\ref{modeloV2Final}), see also (\ref{modeloV2}). 
The dotted line
was obtained using the same values of the parameters as the dash-dotted line
but neglecting the contribution of the second term at the right hand side of 
(\ref{modeloV2}). The difference emphasizes the fact that the new effect may 
be important for correctly accounting for the data at large temperatures. 
The values of the parameters are given in Table \ref{table1}
and in the text. The value of the surface energy for the case of the 
dash double-dotted line was $\Gamma_0=4.5gs^{-2}$. }
\label{V}
\end{figure}

The behavior of the average growth rate for lysozyme aggregation in terms of
temperature at constant supersaturation was measured in Refs. \cite{pusey2} and \cite{pusey}. 
The experimental data (symbols) are shown in Fig. \ref{V} 
together with the results (lines) given by the expression
\begin{equation}\label{modeloV2}
V_a = - \frac{D}{k_BT} \left\{ \left[\frac{\partial }{\partial R}\Delta G(R(t),T)\right]_T 
- \left[\frac{\partial }{\partial T}\Delta G(R(t),T)\right]_R \frac{3}{\alpha_T R} \right\} ,
\end{equation}
obtained from our model after substitutiing Eq. (\ref{ST-Req}) into (\ref{R-dot})
and invoking the definition of the thermal expansion coefficient 
$\alpha_T (T)\equiv (1/V)\left[\partial V/ \partial T\right]_p$, which in 
the present case leads to the relation
$\alpha_T  = (3/R)\left[\partial R/ \partial T\right]_p$, \cite{ragone}. 

This model relation for the average growth rate involves the competition of two
forces coming from the dependence of the Gibbs energy on the radius and the temperature. 
The first term is the usual isothermal contribution whereas 
the second term comes from the temperature dependence of the Gibbs energy 
and is therefore influenced by variations the surface entropy $S_s$ and the thermal 
expansion coefficient $\alpha_T$. 
The results obtained lead us to conclude that a consistent description of the process necessarily entails to consider the coupling of the growth kinetics with the temperature through the second term in Eq. (\ref{modeloV2}).

In order to reproduce the experimental data, we have assumed that near the 
characteristic temperature $T_0$ the thermal expansion coefficient can be expanded 
in the form $\alpha_T \simeq \alpha_0 + \delta \left[{(T-T_0})/{T_0}\right]^a$, 
with $\alpha_0 $ a reference value of the thermal expansion coefficient. 
The scaling prefactor $\delta$ is related to the derivative of the thermal expansion
coefficient with respect to the temperature and $a$ is a scaling exponent. 
From this expression for $\alpha_T$ one can therefore 
infer that the dependence of the average crystal/aggregate radius on temperature
is of the form $R(T)=R_0\, exp[{\cal F}(T)]$ with $R_0$ the 
value of the radius at $T_0$ and ${\cal F}(T)=\int_{T_0}^T\,\alpha_T dT$.
For simplicity in the comparison with experiments \cite{pusey}, we have assumed the 
following linearized expression
for the radius: 
$R(T)\simeq R_0\left[1+\alpha_0 (T-T_0)+\delta T_0^{1+\epsilon}(T-T_0)^{-\epsilon} \right]$ 
where $\epsilon=1+a$. Finally, assuming that in the range of temperatures considered 
the phenomenological coefficient $D$ is constant, we obtain the formula
\begin{eqnarray}
\label{modeloV2Final}
V_a = - \frac{D}{k_BT} \left\{ 8 \pi \left[\Gamma_0+\Delta T \left(b\Delta T-S_0^{(s)}\right)-\frac{3}{2}\frac{\left(2b\Delta T-S_0^{(s)}\right)}{\alpha_T(T)}\right]  R(T)\right.
\nonumber \\
\,\,\,\,\,\,\,\,\,\,\,\,\,\,\,\,\,\,\,\,\,\,\,\,\,\,\,\,\,\,\,\,\,\,\,\,\,\,\,\,\,\,\,\,\,\,\,\,\,\,\,\,\,\,\,\,\,\,\,\,\,\,\,\,\,\,\,\,\,\,\,\,\,\,\,\,\,\,\,
\left. -\frac{4\pi}{\nu_p}\left(1-\frac{1}{T\alpha}\right) kT \ln\left|\frac{{\cal C }}{c_i}\right|  R(T)^2 \right\} ,
\end{eqnarray}
where we have defined $\Delta T \equiv T-T_0$.

The values of the parameters considered in the cases studied are the following: 
$R_0=0.1 \mu m$, $\nu_p = 0.049\mu m$,$\delta=0.123$ and $\epsilon=0.37$. 
The comparison was done using the value of the thermal expansion coefficient of
lysozymes reported in Ref. \cite{kurinov}: $\alpha_0 =1.35\cdot10^{-4}K^{-1}$, 
valid whithin the temperature range: $298-323\,K$. The 
value obtained for the surface energy: $\Gamma_0=5.975gs^{-2}$ is compatible
with that of the surface tension in a lysozyme solution reported in  
Ref. \cite{turco}: $\gamma \simeq 35gs^{-2}$.  For the
diffusion coefficient we have obtained $D \simeq 1\cdot 10^{-10}\mu m^2 s^{-1}$.
This apparently small value may be understood by considering that the
diffusion coefficient in solids can be expressed as \cite{ragone}: 
$D=D_0\,exp(-E_{ac}/k_BT)$. Here $D_0$ is a reference value of the diffusivity and 
the so called exponential pre-factor that depends on the characteristic energy $E_{c}$.
Taking into account that $D_0 \simeq 2\cdot10^{-2}\mu m^2 s^{-1}$, see 
Ref. \cite{velev}, and assuming that $T \sim 280 K$, then we may estimate 
that $E_{c}\simeq 19 k_BT$. This result is compatible with the value of the 
adsorption energy reported in Ref. \cite{velev}, $E_{ads}\simeq 12k_BT$ (see also Ref.\cite{pusey2}).
Finally, the values of the remaining parameters are given in Table \ref{table1}.
\begin{table*}[tbp]
\centering
\begin{tabular}{|c|c|c|c|c|c|c|}
\hline
  &  ${4\pi} \Delta \tilde{\mu}/{3 v_p}$& $S_0^{(s)} $ &${4\pi} \Delta \tilde{\mu}/{3 v_p}\Gamma_0$& $\Gamma_0/S_0^{(s)} $ &$b$  \\
\hline
Line&$g/\mu m s^2$ & $g /s^{2}K$ &$1/\mu m$& $K $& $-$   \\
\hline
  Solid           &   0.20      & 16.60  &0.034& 0.34 &  $0.98$     \\
  Dashed        &   0.18     & 16.00    &0.031& 0.37 & $0.98$      \\
  Dash-dotted &   0.16    &  12.50   &0.026& 0.48& $0.85$       \\
  Dotted         &    0.06    &  5.75 &0.009& 1.04 & $0.40$     \\
\hline
\end{tabular}
\caption{Values of the parameters entering  Eq. (\ref{modeloV2Final}) as obtained by fitting the experimental data shown of Fig. \ref{V} which are in aggreement with those given in Ref.  \cite{pusey2} and \cite{kurinov}.
}
\label{table1}
\end{table*}

From the results obtained, we can conclude that the maximum of the average growth
rate $V_a(T)$ is due to the presence of a minimum of the Gibbs energy and therefore
a consequence of the competition between volumetric and surface effects. 
For a fixed value of $D$, the initial steep and the amplitude of the maximum 
of $V_a$ are associated respectively to the supersaturation parameter and to the 
surface entropy of the crystal/aggregate, that dominates the surface energy at 
low temperatures, this fact is shown in the first and second columns of Table \ref{table1}. 
The second term of the surface energy controls in turn the decrease of $V_a$ 
for larger values of the temperature. For smaller values of the parameter $b$ 
(having the dimensions of a heat capacity per unit area), the tail of the curve becomes 
larger thus improving the fit, see Fig. 3 and Table \ref{table1}.
The decay of $V_a$ is also affected by the value and dependence on temperature of the
thermal expansion coefficient $\alpha_T$. This fact is illustrated in our main finding, Fig. 3,
through the dash-dotted and dotted lines. The dash-dotted line 
was obtained by considering the non-isothermal correction from expression 
(\ref{modeloV2}) or (\ref{modeloV2Final}), whereas the dotted line 
is the result in the isothermal case when the free energy of Eq. (\ref{modeloV2}) 
does not depends on temperature.
It is convenient to emphasize that the contribution coming from the Soret-type effect 
may be relevant in explaining experimental results as those discussed here, because the errors reported
for the measurements of $V_a$ (not shown here) are significantly larger for 
intermediate values of $T$ (i.e., around the maximum) whereas they were small 
for the high temperature regime, see Ref. \cite{pusey}.  A similar behavior
of the average growth rate in terms of the temperature has been also reported
from simulations in Ref. \cite{jacek2006}.

At constant supersaturation, the results for the ratio 
${4\pi} \Delta \tilde{\mu}/{3 v_p}\Gamma_0$ shown in the third column of 
Table \ref{table1} indicate that when the surface energy $\Gamma_0$ of the 
crystal/aggregate is smaller (solid and dashed lines) the maximum of $V_a(T)$ appears 
at temperatures larger than those corresponding to the case of smaller surface energies,
in which the maximum decreases in amplitude and shifts 
towards the region of lower temperatures (dash-dotted and dotted lines). 
Surface energy inhibits the formation of the crystal/aggregate whereas surface 
entropy favours it, see the fourth column in Table \ref{table1}.

\begin{figure}[tbp]
\mbox{\resizebox*{15.0cm}{7.0cm}{\includegraphics{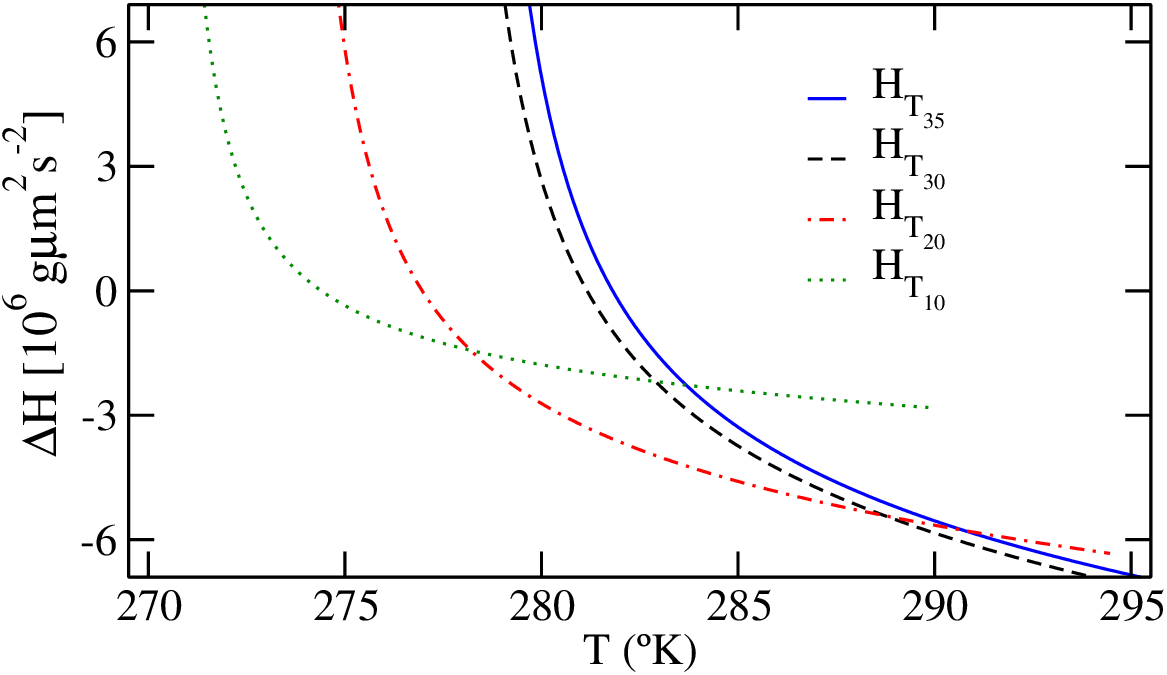}}} {}
\caption{Standard crystallization enthalpy $\Delta H(T)$ as a function
of temperature coming from Eqs. (\ref{GCNT}) and
(\ref{Gibbs-Helm}) for the values of the parameters obtained from the fitting 
shown in Fig. \ref{V}.
}
\label{Enthalpy}
\end{figure}

Another relevant and measurable quantity \cite{pvekil,navarro} that can be obtained from our model is the enthalpy change with temperature. The standard crystallization enthalpy change denoted by $\Delta H$ can be determined from the Gibbs-Helmholtz equation. 
The sign of $\Delta H$ indicating endothermic ($\Delta H>0$) or exothermic ($\Delta H<0$) crystallization can be obtained from \cite{navarro,pvekil} 
\begin{equation}\label{Gibbs-Helm}
\Delta H=-T^2\left(\frac{\partial}{\partial T}\frac{\Delta G}{{T}}\right)_p.
\end{equation}
Making use of Eqs. (\ref{GCNT}), (\ref{GammadeT}) and (\ref{Gibbs-Helm}) with the values of the parameters obtained in the fit of the experimental data, we obtain the behavior of the
enthalpy in terms of temperature shown in Fig. \ref{Enthalpy}. As in the case of the Soret-type coefficient, the change of sign of the enthalpy coincides with the presence of a minimum of the free energy shown in Fig. 1. The quantity $\Delta H$ 
was calculated in \cite{pvekil} for the case of oxidized-and-mutant hemoglobin C crystal 
formation.

\section{Conclusions}

We have analyzed the influence of temperature variations on protein crystallization,
a phenomenon significant for example in biomedicine where small changes in temperature from one tissue 
to another or in the same tissue with time may have an important influence on the crystallization 
of proteins that occur in living systems \cite{Bartling,clinic}.

In particular, we have presented a kinetic model which considers the dependence of the Gibbs free 
energy on temperature to calculate the average growth rate of lysozyme crystals.  
An analysis of the mesoscopic entropy production of the growth process reveals 
the presence of two contributions to the growth rate. The first one is similar to 
that for isothermal growth and can be computed from classical nucleation theory; 
the second stemming from the coupling between growth, viewed at short time 
scales as diffusion along the size of the cluster coordinate, and temperature differences. 
This second contribution constitutes a novel Soret-type effect. As we have shown, this effect is essential to correctly explain experimental results for the average growth rate of crystals/aggreagates in terms of the temperature. 

We used our model to reproduce experiments performed at constant supersaturation
in the high temperature range. The quantitative agreement between experiments and 
our model formula (\ref{modeloV2Final}) is very good even at higher temperatures, 
that are difficult to capture with a single expression for the free energy.
Other models based on the kinetic roughening hypothesis also reproduce the observed crossover of the growth rate \cite{pusey2,liu}. The obtained fit for this quantity is good at low temperatures but at high temperatures it deviates substantially from the measurements. In this domain, the fit provided by our model is better. The fact of reproducing the growth rate in the whole temperature range allows for a systematic interpretation of the experiments in terms of the surface energy and entropy of the crystal/aggregate. Our results also show the important role played by the thermal expansion coefficient in determining the behavior of the average growth rate with temperature.
 
The model can be generalized to describe non-spherical, polygonal-type
crystalline nuclei by explicitly considering the local curvatures $R_i^{-1}$ of their $i$-faces. 
In this case, each face will show different behaviors of the corresponding 
surface energies $\Gamma_{0,i}$, entropies $S_{s,i}$ and components of thermal expansion tensor $\alpha_{T_i}$ leading to different behaviors of the corresponding growth rates $V_{a,i}(T)$, \cite{pusey1,pusey2,pusey,chern,oxtoby}.

The present study may contribute to a better understanding of several 
experimental efforts \cite {pusey2,ducrg,navarro,pvekil} examining the dependence of pure crystal/aggregate 
morphologies with temperature. Future examination of 
the growth process involving nucleation promoters such as salts is left for 
another study \cite{ducrg,poon}.

\section*{Acknowledgements}
The authors are indebted to Prof. R. Winkler, Dr. M. Ripoll, Prof. S. Wiegand, Prof. J. Dhont from FZ J{\"u}lich, and to Mrs N. Kruszewska and Dr. J. Si\'odmiak (Bydgoszcz) for illuminating discussions and useful remarks. AG wishes to thank Prof. R. Winkler for short-term hospitality in FZ J{\"u}lich as of the beginning of year 2010. A Polish-Slovenian common task on nanostructured solutions MNiSW-DSM-WWM-183-1881-12/MS/10 is to be mentioned in parallel with BS5/2009 internal-grant support of the domestic institution of AG. The task has been undertaken during the visiting-professorship scheme in Bydgoszcz of ISH, with partial financial support by DGAPA-UNAM under the PASPA program and grant No. IN102609.
Finally, this work has also been supported by the MICINN of the Spanish Government Grant No. FIS2008-01299.

\end{document}